\documentclass[9pt,twocolumn,twoside]{pnas-new}

\templatetype{pnasresearcharticle}

\usepackage{caption}
\usepackage{subcaption}

\newcommand{\onlinecite}[1]{\hspace{-1 ex} \nocite{#1}\citenum{#1}}

\title{From data to noise to data: mixing physics across temperatures with generative artificial intelligence}

\author[a,b]{Yihang Wang}
\author[a,b]{Lukas Herron} 
\author[b,c,1]{Pratyush Tiwary}

\affil[a]{Biophysics Program, University of Maryland, College Park, MD 20742}
\affil[b]{Institute for Physical Science and Technology, University of Maryland, College Park, MD 20742}
\affil[c]{Department of Chemistry and Biochemistry, University of Maryland, College Park, MD 20742}

\leadauthor{Wang} 

\significancestatement{While it is tempting to use high temperature simulations to infer observations about low temperature, it is not always clear how to do so.  Here we demonstrate how using generative artificial intelligence we can mix information from simulations conducted at a set of temperatures, and generate molecular configurations at any temperature of interest including temperatures at which simulations were never performed. The configurations we generate carry correct Boltzmann weights, and our model minimizes the generation of spurious unphysical configurations. We demonstrate its use here through combining with replica exchange molecular dynamics in a post-processing framework for sampling peptide and ribonucleic acid. We believe the framework is extensible to generic simulations and experiments for mixing control parameters other than temperature.}

\authorcontributions{PT, YW designed research. YW, LH performed research. PT, YW, LH wrote the manuscript.}
\authordeclaration{There are no competing interests to declare.}
\correspondingauthor{\textsuperscript{1}To whom correspondence should be addressed. E-mail: ptiwary@umd.edu}

\keywords{Molecular Simulations$|$ Generative Artificial Intelligence$|$ Enhanced Sampling} 

\begin{abstract}
Using simulations or experiments performed at some set of temperatures to learn about the physics or chemistry at some other arbitrary temperature is a problem of immense practical and theoretical relevance. Here we develop a framework based on statistical mechanics and generative Artificial Intelligence that allows solving this problem. Specifically, we work with denoising diffusion probabilistic models, and show how these models in combination with replica exchange molecular dynamics achieve superior sampling of the biomolecular energy landscape at temperatures that were never even simulated without assuming any particular slow degrees of freedom. The key idea is to treat the temperature as a fluctuating random variable and not a control parameter as is usually done. This allows us to directly sample from the joint probability distribution in configuration and temperature space. The results here are demonstrated for a chirally symmetric peptide and single-strand ribonucleic acid  undergoing conformational transitions in all-atom water. We demonstrate how we can discover transition states and metastable states that were previously unseen at the temperature of interest, and even bypass the need to perform further simulations for wide range of temperatures. At the same time, any unphysical states are easily identifiable through very low Boltzmann weights. The procedure while shown here for a class of molecular simulations should be more generally applicable to mixing information across simulations and experiments with varying control parameters.
\end{abstract}

\dates{This manuscript was compiled on \today}

\begin{document}

\maketitle
\thispagestyle{firststyle}
\ifthenelse{\boolean{shortarticle}}{\ifthenelse{\boolean{singlecolumn}}{\abscontentformatted}{\abscontent}}{}


\dropcap{H}ow  does  one  learn  physics  and  chemistry at  a  certain  temperature  on  the  basis  of  experiment  or  experiments  performed  at  certain  other  temperatures? The Arrhenius  Equation\cite{laidler1984development} provides the  simplest  way  to  do,  if  one  is  willing  to make significant simplifying assumptions about the energy landscape  of  the  system. Often arbitrarily complex systems indeed display kinetics conforming to the Arrhenius picture\cite{scalley1997protein}, but frequently one also observes its violations\cite{chan1998protein}. Even when the Arrhenius picture is applicable, it only allows extrapolation of a single number i.e. reaction rate across the temperatures. It thus remains desirable to develop techniques allow inferring much more detailed information across a temperature range. These could be thermodynamic observables such as relative probabilities of a molecule's different conformations or of a crystal's different polymorphs, or more detailed kinetic observables than just an overall rate constant. Even more generally, given observations of the positions and velocities of all constituents of a $N-$body system at some set of temperatures, we seek to estimate what they will be at a temperature where the experiment or simulation was never even performed. The problem is hard because the probability distribution connecting configurational coordinates across temperatures is hard to sample from, especially since $N$ is extremely large for most systems of interest, equaling at least tens of thousands.

Here we propose a framework using generative Artificial Intelligence (AI) that learns to efficiently sample such high-dimensional, complex probability distributions for $N-$body molecular systems valid across temperatures. There are two central ideas guiding our framework. Firstly, we do not treat the temperature $T$ as just a control parameter. By noting that the temperature is a measure of the average kinetic energy, we instead work directly with the fluctuating kinetic energy, using equipartition theorem to define an instantaneous, effective temperature which we still call $T$. For finite $N$ not yet in the thermodynamic limit, $T$ so defined will display significant fluctuations proportional to ${1 / \sqrt{N}}$. Secondly, if we have experiments or simulations performed at temperatures $T_1, T_2,...T_K$, we can view them all together as being sampled from the same, but unknown, joint probability $p(\mathbf{x},T)$ for $N-$body configurations $\mathbf{x}$. We then use a generative AI method, specifically denoising diffusion probabilistic models (DDPM)\cite{DPM,denoising} to generate many more samples from $p(\mathbf{x},T)$, given the spare, high-dimensional dataset.

To learn a $p(\mathbf{x},T)$ from such a high-dimensional and sparse dataset, we use the denoising diffusion probabilistic model (DDPM),\cite{DPM,denoising} a generative artificial intelligence (AI) framework inspired by non-equilibrium dynamics to approximate and sample from very high-dimensional distributions. By learning $p(\mathbf{x},T)$ we are referring to the ability to generate many more samples from $p(\mathbf{x},T)$ given the data set we have from simulations or experiments performed so far.  DDPM has been shown to possess the ability to infer and learn the underlying relationship from complicated data i.e. high-dimensional and with complex correlations, but also noisy and sparse\cite{DPM,denoising}. DDPMs learn two diffusion processes expressed through stochastic deep neural networks, which are called noising and denoising. The forward diffusion or noising converts the samples from high-dimensional, structured and unknown probability distribution into simpler and analytically tractable white noise. The backward diffusion or denoising learns the mapping back from noise to meaningful data. The central idea is that it is easy to generate numerous samples from the noisy distribution, which can then be denoised back to structured data. The process has been demonstrated to be comparable or even outperform other generative AI models for generating high quality samples, and in its ability to model in an unsupervised way the underlying semantics behind meaningful data.\cite{DPM, denoising}

In spite of their promising potential, to the best of our knowledge DDPMs have not yet been used in the context of mixing information across processing conditions as done here. We are able to achieve this here through recognizing that for finite-size systems, as one has in molecular simulations, the temperature can be associated with a random variable that has significant fluctuations and not just a control parameter. 
The protocol developed here should be applicable quite generally to data coming from simulations or experiments at different temperatures. It should also be extensible to mixing data from control parameters other than temperatures - possibly including concentration, pressure and volume. Here we focus on a specific class of molecular simulations known as Replica exchange molecular dynamics (REMD)\cite{REMD,hansmann1997parallel} that use information generated at a ladder of temperatures to swap configurations across temperatures. REMD has been extremely powerful over the decades for the study of molecular systems with rough energy landscapes for fundamental science and practical applications.\cite{abrams2014enhanced,abel2017advancing} Numerous advances have been introduced over this basic idea in order to make it more efficient computationally\cite{REST,REST2,ballard2009replica,trebst2006optimized,nadler2007optimizing,kim2010generalized,chodera2011replica,gil2015enhanced} and it continues to be an area of very active research. We demonstrate how DDPM applied to all-atom, femtosecond resolution REMD simulations of a small peptide and a RNA nucleotide very significantly improve the quality of data generated across temperatures, including temperatures at which the simulation was never performed and even at temperatures outside the ladder of temperatures. significantly improves the estimates of free energies made through REMD and providing accurate sampling in parts of configuration that were not previously visited in the lowest temperature replica during REMD. This could be metastable states or transition states. We also show how this can be used to generate samples at temperatures not included in the ladder of replicas. The samples we generate have thermodynamic relevance and correspond to correct Boltzmann weights as opposed to being just wild hallucinations.\cite{denoising,deep_learning_book} Due to its generalizable framework and simple post-processing nature of application we thus believe that this work should be extensible to a wide range of studies to supplement and extend the range of simulations/experiments without actually performing them at all conditions of interest.

\section*{Results}

\textbf{Defining a fluctuating effective temperature.} We work with the equivalent $p(\mathbf{x},\beta)$ where inverse temperature $\beta = {1 \over k_B T}$ and $k_B$ is Boltzmann's constant. Molecular dynamics (MD) or Monte Carlo (MC) methods allow sampling configurations $\mathbf{x}$ as per the equilibrium probability $p_{\beta}(\mathbf{x}) \equiv {e^{-\beta U(\mathbf{x})} /  Z}$, where $U(\mathbf{x})$ is the potential energy of the $N-$body system and $Z = \int d\mathbf{x} e^{-\beta U(\mathbf{x})}$ is the partition function. For systems of practical interest in biology, chemistry and materials science,  if $T$ is not large enough, it becomes nearly impossible to sample reliably from $p_{\beta}(\mathbf{x})$ as many regions of interest in configuration space will have $e^{-\beta U(\mathbf{x})} \approx 0$. To deal with this problem, in REMD one simulates $K$+1 replicas of the system at temperatures $\beta = \beta_0 > \beta_1 > ... > \beta_K $. For low enough $\beta_K $ or equivalently high enough temperature $T_K$, the sampling from $p_{\beta_K}(\mathbf{x}) \equiv {e^{-\beta_K U(\mathbf{x})} /  Z_K}$ is expected to be more ergodic. One then periodically exchanges conformations between consecutive pairs of replica with a Metropolis-type acceptance probability that depends on the potential energies of the two replicas and their temperatures. This way even the low-temperature $\beta_0$ replica can explore configurations that it would not have otherwise visited. 

We now argue that by treating the temperature just as a control parameter REMD is not making full use of the information gathered across the ladder of temperatures. In  most current incarnations of REMD (excluding exceptions such as Ref. \cite{gallicchio2005temperature}), all that the higher temperatures do is to help the lower temperature replicas discontinuously appear in different locations of the configuration space. We take a slightly different view of the temperature, in REMD here but easily generalizable to other thermodynamic variables which show fluctuations for finite system size. By working with an approximate, effective temperature, we treat it as a random variable instead of a control parameter. To do so we work with an effective, instantaneous temperature of the system associated with its kinetic energy instead of the temperature of the heat bath that we expect the thermostat to enforce on average. More rigorously thus, we are sampling the joint $p(\mathbf{x},\kappa)$ where the per-particle kinetic energy $\kappa$ is related through its ensemble average to the temperature, i.e. $\langle \kappa \rangle = {3 \over 2\beta} $. We emphasize that our effective temperature is not the true thermodynamic temperature - however for the sake of simplicity we still use the symbol $T$ or $\beta$ to denote this effective temperature or its inverse respectively.

Our motivation in doing so is that all replicas across different temperatures can be viewed as being sampled from the same joint probability $p(\mathbf{x},\beta)$ -- as opposed to different replicas sampling from respective $p_{\beta_K}(\mathbf{x})$. This change of perspective allows us to combine together the data collected from different temperature as having arisen from the same, although intractble, probability distribution $p(\mathbf{x},\beta)$. 

\textbf{Challenge in sampling the intractable joint probability across configurations and temperatures.} Our task at hand now is to learn the joint probability $p(\mathbf{x},\beta)$ given sampling that has already been performed in REMD across $\mathbf{x}-$space and temperatures $\beta = \beta_0 > \beta_1 > ... > \beta_K $. There are two main challenges in this. The first challenge comes from the curse of dimensionality: the memory or computational resources needed to track a very high-dimensional distribution function increase exponentially as the number of degrees of freedom increases. For instance, for REMD of a small 9-residue peptide in explicit water, which we study here, we exchange all 4749 atomic coordinates between the replicas, but for the purpose of analysis we set $\mathbf{x}$ as 18 Ramachandran dihedral angles. This means we already have a 19-dimensional space where binning procedures are out of the question. The second challenge comes from the sparsity of the data. Most of our samples come from high probability regions $p(\mathbf{x},\beta)$ and we have very few samples for low-probability states, or for high-probability states at low temperature due to inefficient exchange betweem replica. In summary we have sparse sampling of data points in very high-dimensional $(\mathbf{x},\beta)-$space and wish to construct $p(\mathbf{x},\beta)$ from this information so that we can create many more samples at any temperature of interest.

\begin{figure*}[hbt]
  \centering
  \includegraphics[width=1\textwidth]{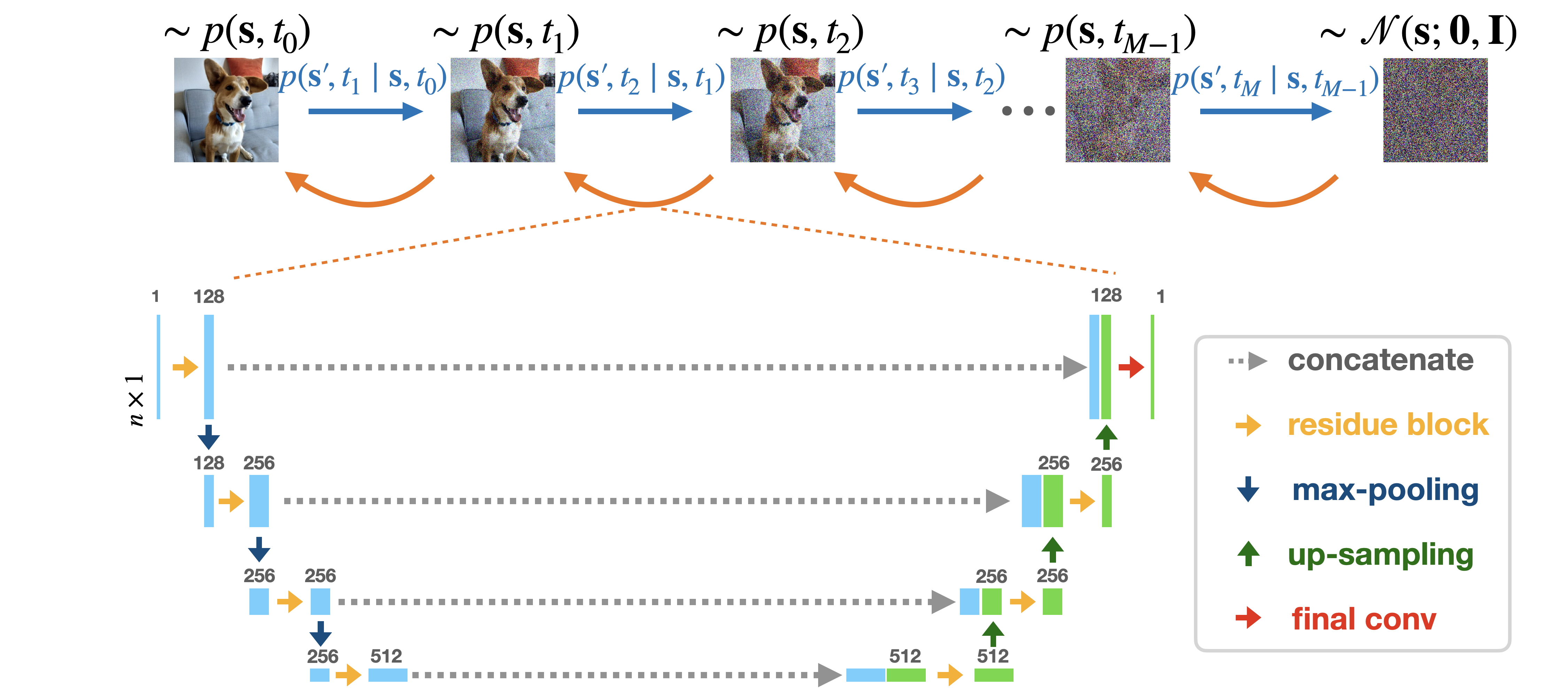}
  \caption{ Lower panel shows the neural network architecture used in this work. It has the basic structure of the U-Net model.\cite{u_net} During the diffusion process, which is indicated by the direction of blue arrows in the upper panel, noise is gradually added to the sampled data, in this case the picture of a very good boy, through a diffusion process labeled with diffusion step $t_i$. This changes the sampled distribution (for example, more pictures of dogs) to a simpler isotropic Gaussian distribution from which one can easily generate more samples. An AI model is then trained to reverse such a diffusion process and starting from sampled noise, learn to generate images similar to the input image by following the direction indicated by the orange arrows. It takes an 1-d array $\mathbf{s'}$, which is the noisy sample at diffusion step $t_i$, as input and outputs the parameterization of a Gaussian distribution to get the reversed transition kernel $q(\mathbf{s}, t_{i-1}\mid  \mathbf{s}, t_i)$. Each residue block consists of three components: two 1-d convolution operations with kernel size 3 and a group normalization\cite{group_normalizng} between them. The diffusion step $t_i$ is added to each convolutional block after being transformed by the sinusoidal position embedding\cite{attention}. The final conv label denotes the convolution operation with kernel size equal to 1. Max-pooling reduces size of the features by half, while up-sampling uses the transposed convolution to expand the size of features. }
  \label{fig:U-net}
\end{figure*}


\textbf{Denoising diffusion probabilistic models (DDPM) can generate many more samples from  $p(\mathbf{x},\beta)$.} The main idea behind the use of DDPM here is to learn a simple and easy-to-sample-from distribution $P_{\text{simple}}(\mathbf{x},\beta)$ that approximates the true $p(\mathbf{x},\beta)$. For notation simplicity, we denote $\mathbf{s}=\left\{\mathbf{x},\beta\right \}$ and refer to $P_{\text{simple}}(\mathbf{s})$ henceforth. DDPM does this by learning to reverse a gradual, multi-step noising process that starts with the relatively limited number of samples generated from the distribution $p(\mathbf{s})$ and diffuses to the simpler distribution $P_{\text{simple}}(\mathbf{s})$ that is easy-to-sample. For instance, $P_{\text{simple}}(\mathbf{s})$ could be an isotropic Gaussian. In addition to learning this noising process, DDPMs also learn the reverse denoising process which allows us to go back from samples generated using $P_{\text{simple}}(\mathbf{s})$ to samples that would have been generated from the underlying $P(\mathbf{s})$. Both the noising and denoising processes are modeled using diffusion processes that convert probability distributions to one another, and are implemented using the architecture shown in Fig. \ref{fig:U-net}, which is based on the standard architecture for DDPMs as described in Ref.\cite{denoising}. 
\begin{figure*}[hbt]
  \begin{subfigure}{0.55\textwidth}
  \centering
  \includegraphics[width=0.95\textwidth]{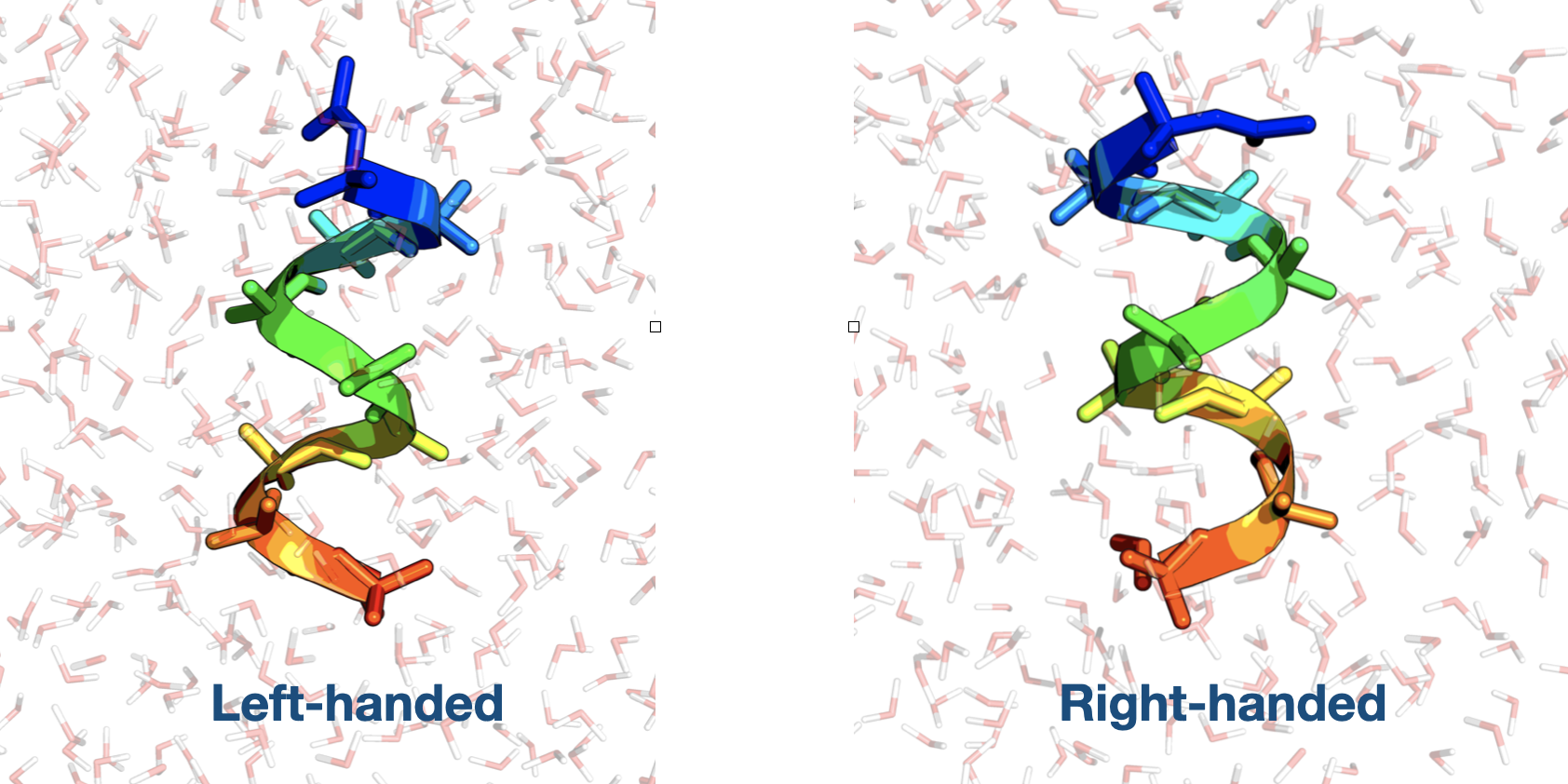}
  \vspace{1.0\baselineskip}
  \caption{ }
  \label{fig:fe_structure}
  \end{subfigure}
  \begin{subfigure}{0.38\textwidth}
  \centering
  \includegraphics[width=0.95\textwidth]{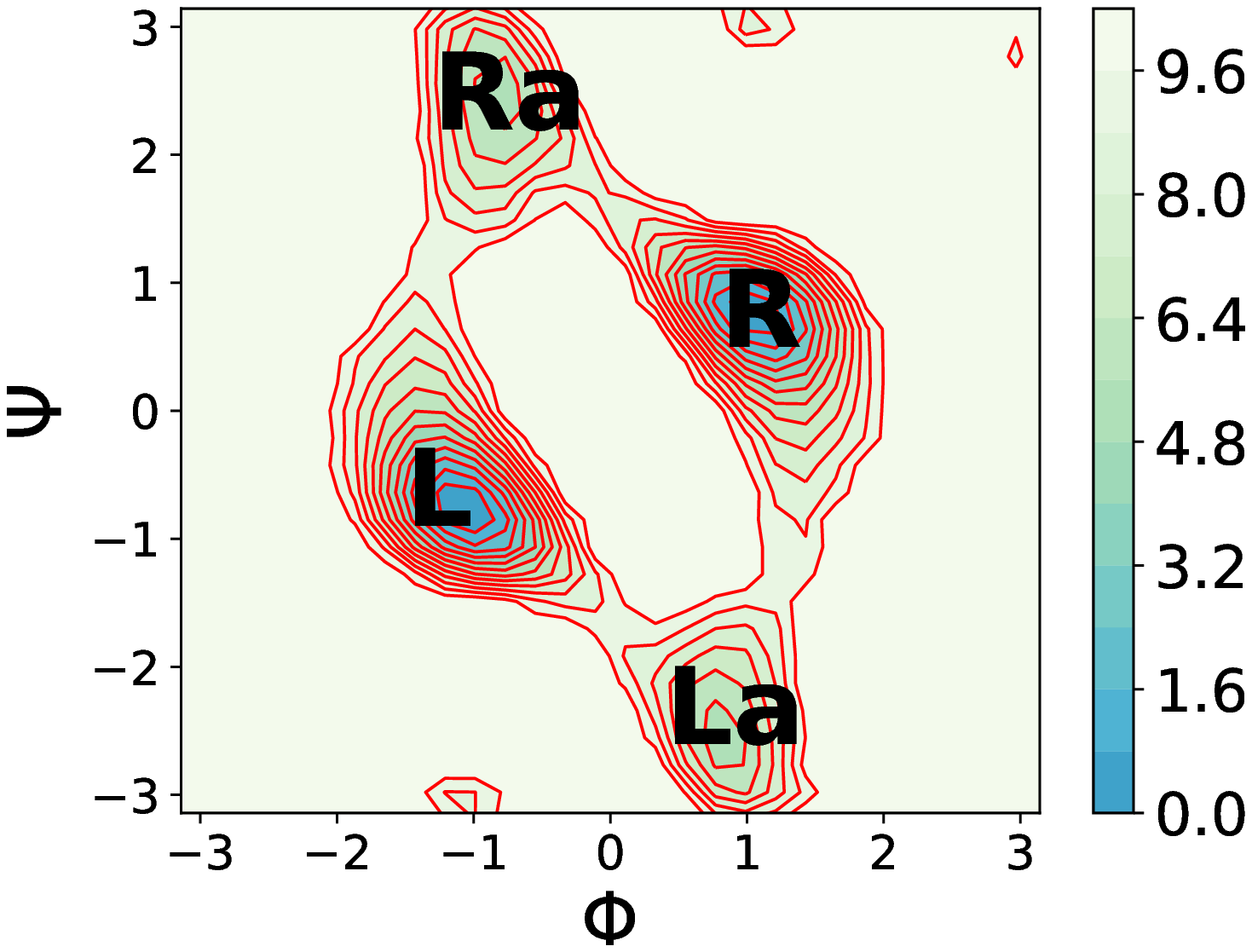}
   \caption{ }
  \label{fig:fe_unbiased}
  \end{subfigure}
  \caption{ (a) AIB$_9$ in left-handed (L) and right-handed (R) conformations in explicit TIP3P water. (b) Free energy profile of residues (with residue 5 as an example) with two stable states labeled as L, R and two excited state labeled as La, Ra. When all residues are in L states, the peptide chain is in left-handed state and similar for right-handed state. Free energies in (b) here as well as all through the manuscript and Supporting Information are provided in units of $k_BT$.} 
\end{figure*}

\textbf{Noising and denoising processes.} The noising diffusion process carried out in the space $\mathbf{s}=\left\{\mathbf{x},\beta\right \}$ that converts $p(\mathbf{x},\beta)$ to the simpler $P_{\text{simple}}(\mathbf{x},\beta)$ can be decomposed into $M$ discrete steps denoted by corresponding transition probabilities $p( \mathbf{s'}, t_{i+1}| \mathbf{s}, t_i)$ where $i\in[0,M]$, $P(\mathbf{s},t_0) \equiv p(\mathbf{x},\beta) $, and $P(\mathbf{s},t_M) \equiv P_{\text{simple}}(\mathbf{x},\beta)$. In DDPM, this noising diffusion process that converts sampled data to essentially noise, is set to be an Ornstein--Uhlenbeck (OU) process in which the transition probability follows an simple Gaussian form. One can then easily generate samples from $P(\mathbf{s},t_0) \equiv p(\mathbf{x},\beta) $. The tricky bit now is to convert these samples back to the original distribution. In Ref. \onlinecite{DPM}, it was shown that this transition reversed diffusion kernel $P( \mathbf{s}, t_{i}| \mathbf{s'}, t_{i+1})$ can also be written in a Gaussian form. A deep neural network (Fig. \ref{fig:U-net}) is then trained using variational inference to learn the approximate reversed transition kernel $Q( \mathbf{s}, t_{i}| \mathbf{s'}, t_{i+1}) \approx P( \mathbf{s}, t_{i}| \mathbf{s'}, t_{i+1}) $. Thus by generating samples from a normal Gaussian distribution, which we can easily do in large numbers, and then passing these through the reversed transition kernel, we can generate samples that follow the target distribution $p(\mathbf{x},\beta)$ as desired. Note that instead of learning the joint probabilities $P(\mathbf{s_1},\mathbf{s_2})$, it can be advantageous to learn the conditional probability $P(\mathbf{s_1}|\mathbf{s_2})$. This can be done through the protocol in Ref. \onlinecite{DPM} by adding a delta function to allow only a subset of $\mathbf{s}$ to change during the nosing and denoising process, i.e. $\delta(\mathbf{s_2} - \mathbf{s'_2})P( \mathbf{s_1}, \mathbf{s_2}, t_{i}| \mathbf{s'_1}, \mathbf{s'_2},  t_{i+1}) $. This is very useful when for instance we are interested in generating samples only at a certain temperature or only in certain regions of the configuration space.
In the most general form of diffusion probabilistic models (DPM),\cite{DPM} the reversed transition kernel $Q( \mathbf{s}, t_{i}| \mathbf{s'}, t_{i+1})$ is considered as $
Q(\mathbf{s}, t-1 \mid \mathbf{s'}, t)=\mathcal{N}(\mathbf{s} ; \tilde{\boldsymbol{\mu}}(\mathbf{s'}, t), \tilde{\boldsymbol{\sigma}}_{t}(\mathbf{s'}, t) )$ and the neural network is trained to learn the mean $\tilde{\boldsymbol{\mu}}(\mathbf{s'},t) $ and variance $\tilde{\boldsymbol{\sigma}}(\mathbf{s'}, t)$. In practice, however, there are many different ways to choose the Gaussian distribution parameterization. In Ref. \cite{denoising}, DDPM was introduced with a new parameterization approach to reduce the complexity of the training task. DDPM got its name because such a design makes the learning task resemble a denoising score matching procedure.\cite{vincent2011connection} In Ref. \onlinecite{denoising}, it was shown that with such a design, DDPM can generate samples of a quality that are comparable or even better than other generative models. 

\textbf{DDPM applied to Replica Exchange MD: Desirables.} We now demonstrate how the above protocol can be applied to mix data collected from different temperatures and configurations in REMD and signficantly improve the quality of sampling. Specifically, we consider the following two challenging tasks:
\begin{enumerate}
    \item Can we improve the sampling quality for the lowest-temperature replica with more accurate probability estimates than directly seen after REMD? This includes being able to generate samples in low probability regions such as transition and metastable states, and reliably estimating their free energies.
    \item Can we generate samples at temperatures that are not even included in the replica ladder? This would include temperatures within the range of the replica ladder and also extrapolation to temperatures outside the range.
\end{enumerate}

\textbf{Peptide conformational transitions.} To demonstrate the performance of DDPM, we first study a small peptide chain Aib$_9$ in explicit TIP3P water\cite{tip3p} using CHARMM36m force-field.\cite{huang2017charmm36m} This 9 residue system (Fig. \ref{fig:fe_structure}) displays rich and complex conformational dynamics\cite{botan2007energy} including the transitions between fully left-handed and right-handed helices. However, even in 4 $\mu s$ unbiased MD at 400K, one can see only around 2--3 transitions between these two dominant equiprobable conformations, and even fewer, if any, transitions to the higher energy metastable states. To improve the sampling of this system, we perform REMD with 10 replicas at geometrically spaced temperatures ranging from 400K to 518K, with temperature increased by 3$\%$ for each replica. The attempt of exchanging configurations was made every 20 ps, with acceptance rate around $1\% \sim 2\%$ between neighboring replicas, which is intentionally kept lower than what one usually has in REMD. This is because we want to show that even in the extreme cases where the number of atoms $N$ is so large that replicas do not have enough overlap, or if one wants to reduce the number of replicas to save computational resources, our DDPM can still do a decent job of complementing REMD and reconstructing the true probability distribution at any temperature of interest. To benchmark our results, we ran unbiased MD at 400 K for $4$ $\mu s$ and at 500 K for $0.6$ $\mu s$.

\begin{figure}[b!]
  \begin{subfigure}{0.5\textwidth}
  \centering
  \includegraphics[width=1.0\linewidth]{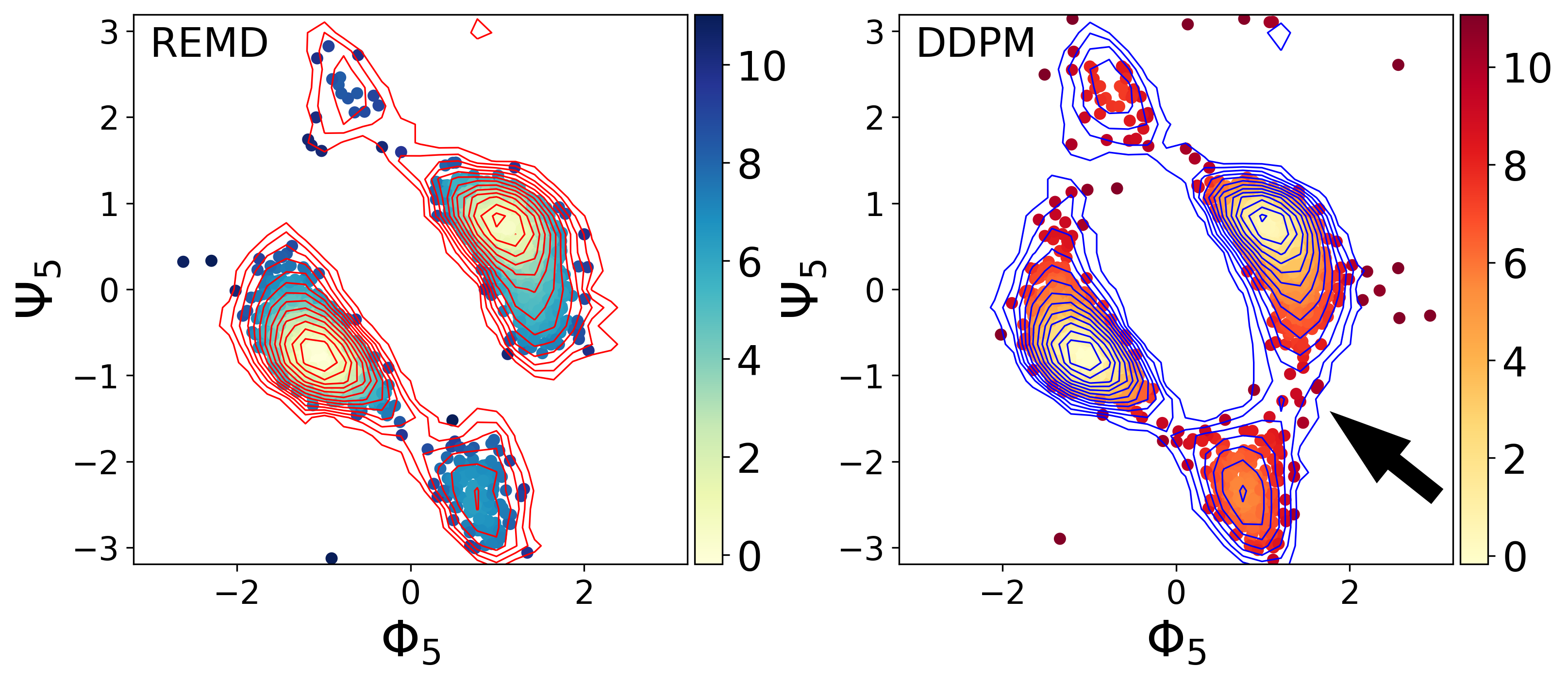}
  \caption{}
  \label{fig:fe_res5}
\end{subfigure}%

\begin{subfigure}{0.5\textwidth}
  \centering
  \includegraphics[width=1.0\linewidth]{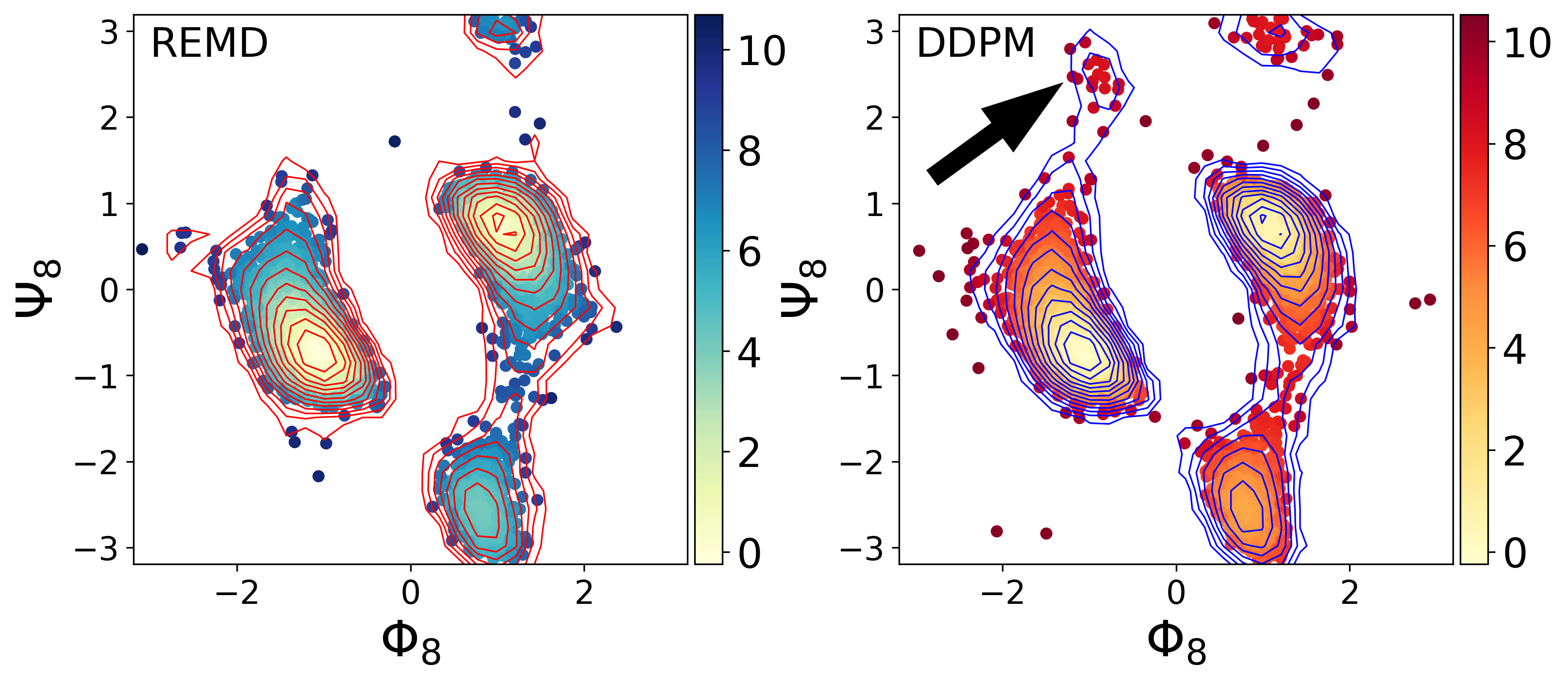}
  \caption{ }
  \label{fig:fe_res8}
\end{subfigure} 
\caption{Samples (dots) from REMD (left) and DDPM (right) at 400K. The Boltzmann weights for different samples are indicated through their free energy (contour lines, separated every 0.74 $k_bT$). (a) Samples generated from DDPM and free energy profile projected on dihedral angles of residue 5.  (b) Samples generated from DDPM and free energy profile projected on dihedral angles of residue 8. DDPM was able to generate samples in states that are not present in the training dataset for both residues, indicated with thick black arrows in the right panels.
}
\label{fig:samples}
\end{figure}



 To quantify the quality of sampling, we focus on the 18 dihedral angles corresponding to all 9 residues $(\Phi_1, \Psi_1, \Phi_2, \Psi_2, \cdots \Phi_9, \Psi_9)$. As shown in the Ramachandran plot in Fig. \ref{fig:fe_unbiased}, the free energy surface along any pair of dihedrals is mainly characterized by four metastable states: two equiprobable low-energy ground states (L, R) and two excited states (La, Ra). The Ramachandran plots for all nine residues in the system look qualitatively similar but the middle residues of the peptide are known to be less flexible with higher energy barriers.\cite{biswas2018metadynamics, SPIB_AIB9} We thus focus on the sampling for residues 5 and 8 as shown in Fig. \ref{fig:samples}. We train our DDPM on a REMD trajectory. At the lowest temperature of interest (400K), this trajectory has not yet achieved sufficient sampling. As shown in Fig. \ref{fig:samples}, DDPM successfully mixes information from all temperatures and configurations, and generates samples in states that are not present in the training dataset for both residues, indicated with thick black arrows in the right panels of Fig. \ref{fig:samples}. Specifically, for residue 5 we can see that the transition states between state R and state La, which were not being sampled in the 400K replica, are populated in samples from DDPM. For residue 8 the improvement is even more striking as the state Ra which was simply not sampled in REMD now gets populated after DDPM. To further quantify the improvement gained due to DDPM, we compare the free energy differences between different configurations from both REMD and DDPM against much longer reference unbiased MD at 400K. As shown in Supplementary Fig. S2, the free energy difference calculated from samples generated by DDPM are much closer to the that from the reference MD. Thus, DDPMs are able to accomplish the first task highlighted above. We also want to highlight that while some spurious samples are generated, seen through dots outside the free energy contours in Fig. \ref{fig:samples}, their Boltzmann weights are very low. Thus, our model ``dreams" new configurations without hallucinating spurious configurations.

\begin{figure*}[tbh]
  \centering
  \includegraphics[width=0.9\textwidth]{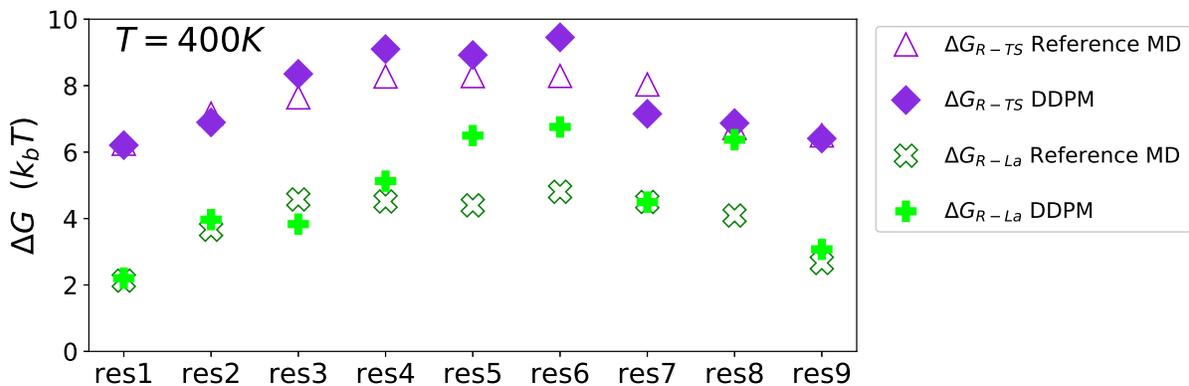}
  \caption{ Free energy differences between different metastbale states $\Delta G$ calculated from samples from DDPM by extrapolating to a temperature at 400 K, which is lower than any temperature included in the dataset. The DDPM model was trained with REMD trajectories with 9 replicas starting at the lowest temperature 412K. Green empty crosses and purple empty triangles are reference free energy differences $\Delta G$ from 4 $\mu s$ long unbiased MD at 400 K. Green filled plus signs and purple diamonds show the energy differences from DDPM. $\Delta G_{\text{L-Ra}}$ is the free energy difference between ground state L ($-2.2<\Phi<-0.1$, $-1.6<\Psi<0.9$) and excited state Ra ($-1.6<\Phi<-0.2$, $1.4<\Psi<3.14$); and $\Delta G_{\text{L-TS}}$ is the free energy difference between the ground state L ($-2.2<\Phi<-0.1$, $-1.6<\Psi<0.9$) and a transition state ($-1.8<\Phi<-0.8$, $0.8<\Psi<1.5$)  } 
  \label{fig:extrapolation}
\end{figure*}

We now move to the second task from the list above, and test DDPM's ability to generate samples by interpolating or even extrapolating across temperatures not considered in the ladder of replicas. In the first example, we use it to generate samples at 500 K, as in the training set with 10 replicas at geometrically spaced temperatures between 400 and 518K, there is no replica with temperature 500 K. As shown in Supplementary Fig. S2b, the $\Delta G$ calculated from samples of DDPM is in good agreement with that from the reference MD at 500 K. In the second example, we completely removed the samples from 400 K replica in the training set and use it train a new model, which is then used to generate samples at 400 K. Even though in the training set, the lowest temperature is 412K, the model can make good prediction of free energy difference between states as shown in Fig. \ref{fig:extrapolation}. We are particularly encouraged by this last finding as in general sampling at lower temperatures tends to be harder than at higher temperatures.

\begin{figure}[hbt]
\begin{subfigure}{0.24\textwidth}
  \centering
  \includegraphics[width=0.95\textwidth]{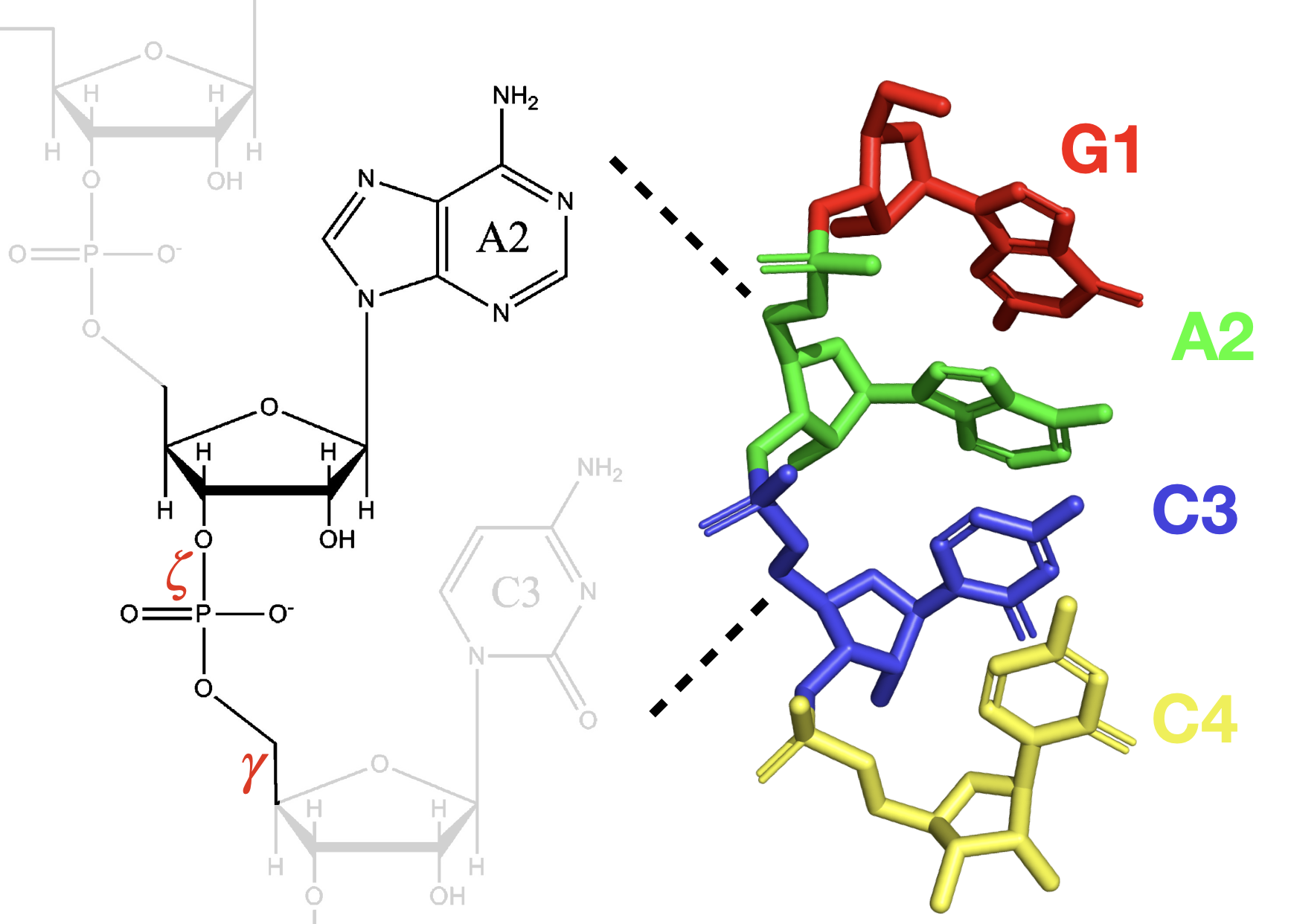}
   \vspace{2.5\baselineskip}
  \caption{  } 
  \label{fig:REMD_structure}
\end{subfigure}%
 \begin{subfigure}{0.24\textwidth}
  \centering
  \includegraphics[width=1.02\textwidth]{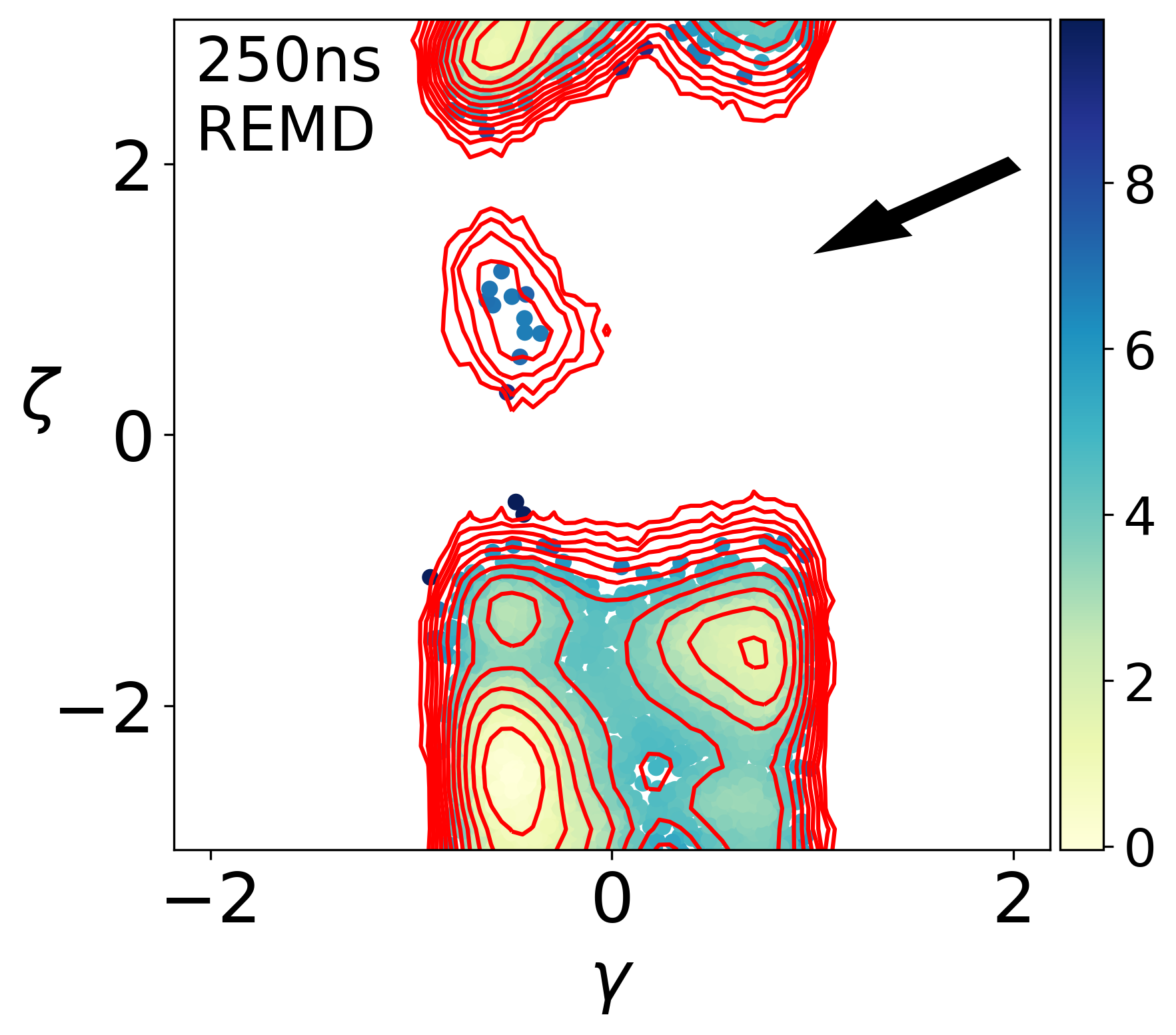}
  \caption{  } 
  \label{fig:REMD_GACC}
\end{subfigure}%

\begin{subfigure}{0.24\textwidth}
  \centering
  \includegraphics[width=1.02\linewidth]{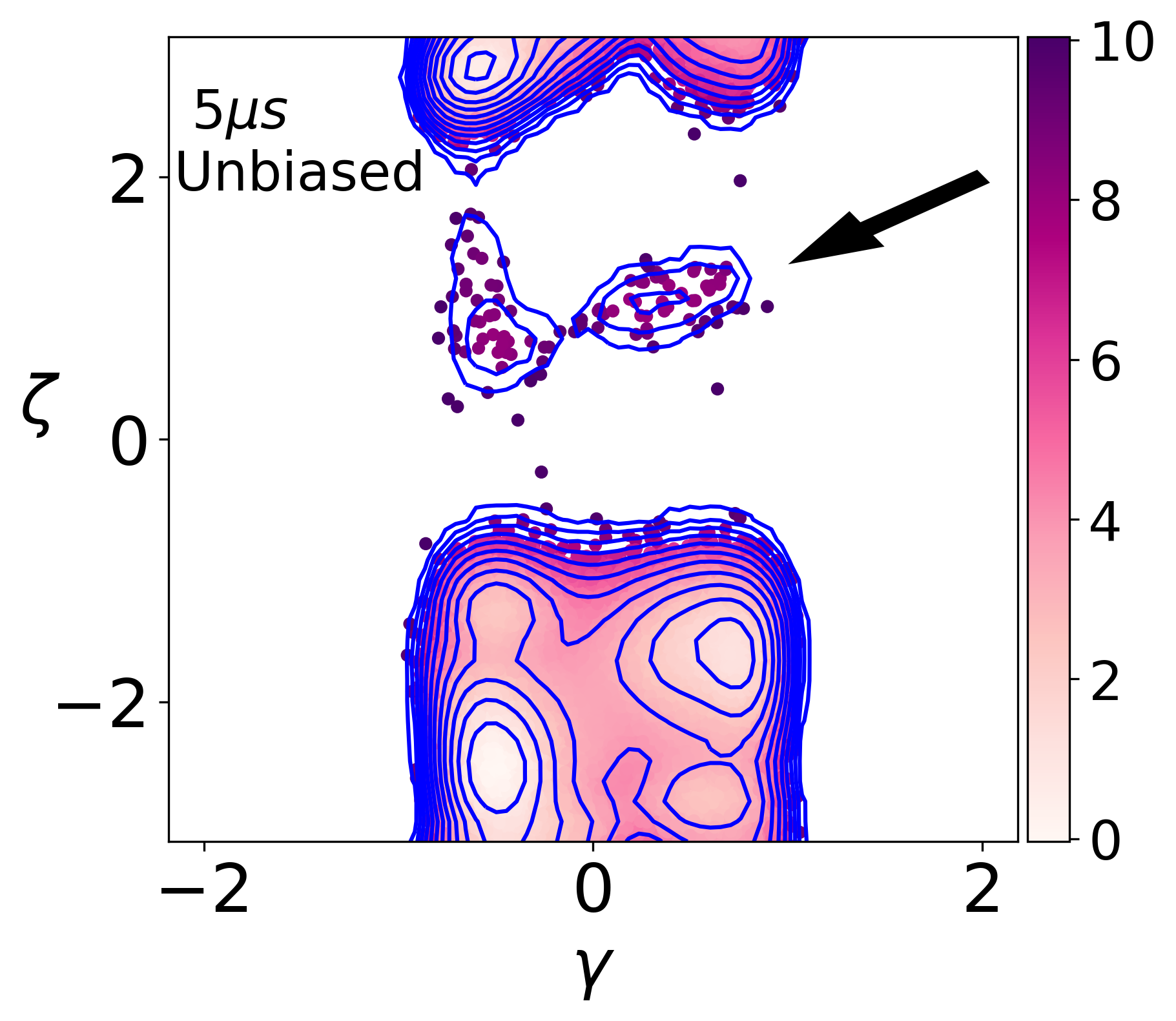}
  \caption{}
  \label{fig:unbiased_GACC}
\end{subfigure}%
\begin{subfigure}{0.24\textwidth}
  \centering
  \includegraphics[width=1.02\linewidth]{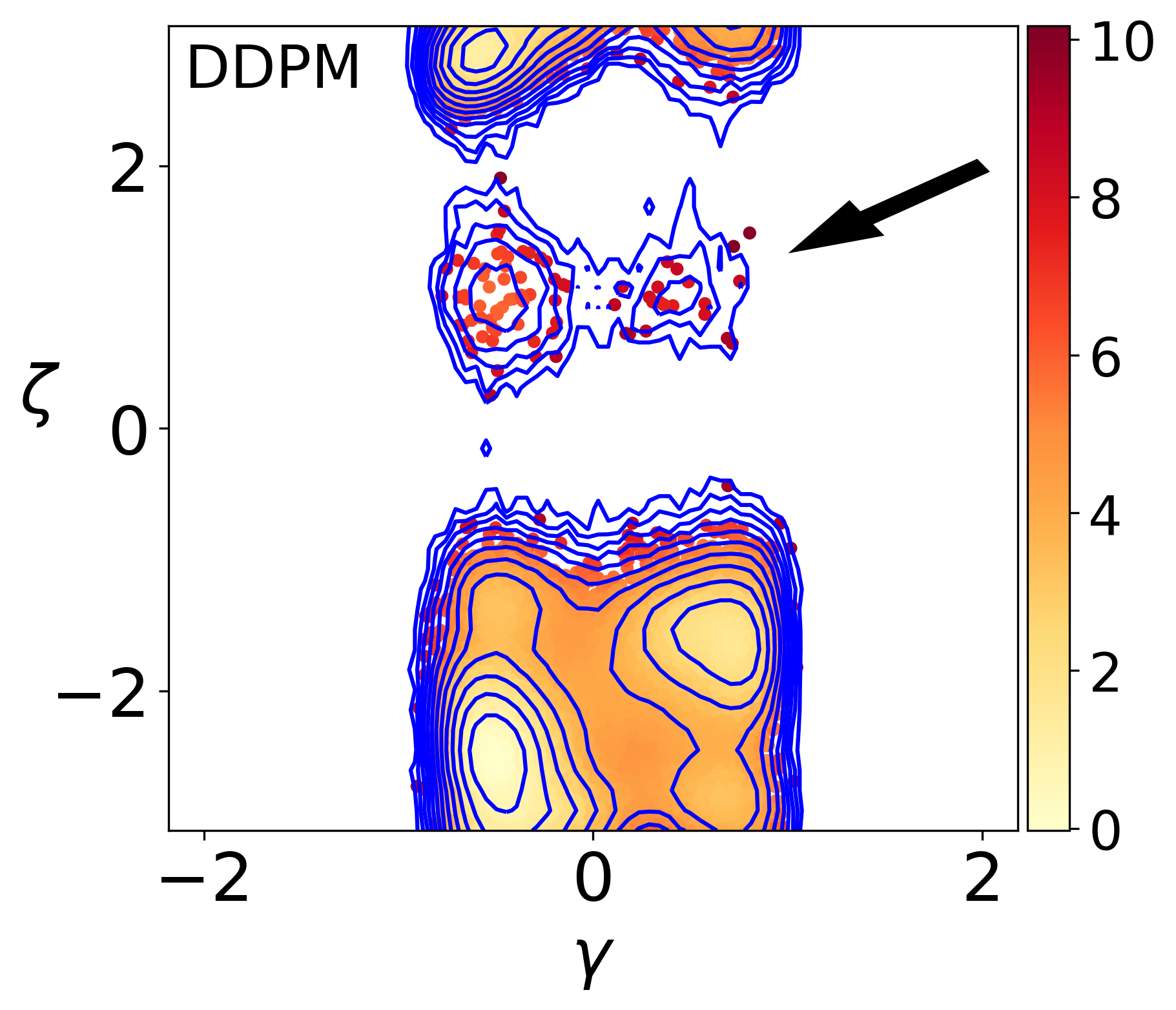}
  \caption{ }
  \label{fig:DDPM_GACC}
\end{subfigure} 
\caption{Projection of samples and free energy profile on dihedral angles $\zeta$ and $\gamma$ of A2 in GACC at 325 K. The Boltzmann weights for different samples are indicated through their free energy (contour lines, separated every 0.74 $k_bT$). (a) The structure of GACC. (b) Samples from REMD. (c) Samples from the benchmark $5 \mu s$ long unbiased MD. (d) Samples from DDPM. The metastable states that are not present in the training dataset are indicated with thick black arrows. }
\label{fig:fes_GACC}
\end{figure}

\textbf{RNA conformational transitions.}
As a second example to illustrate the general applicability of our DDPM+REMD approach, we turn our attention towards sampling RNA conformational ensemble. Rare RNA structures have been previously shown to be biologically relevant \cite{Bannwarth2005, SchulzeGahmen2018, Ganser2019}, but estimating the conformational ensemble still remains computationally intractable using traditional sampling techniques \cite{Sponer2018}. RNA dynamics may occupy a wide range of timescales - from several hours for conformational changes that require breaking base pairs, to picoseconds for more continuous deformations. \cite{Ganser2019} As a consequence, identifying rare transient structures and estimating their contribution to the RNA ensemble has proved to be difficult.
In this example, we consider a GACC tetranucleotide as our model system (Fig. \ref{fig:REMD_structure}). As a single-stranded RNA consisting of four nucleotides labeled G1, A2, C3 and C4, GACC has been previously used as a challenging test system for REMD based sampling methods for its conformational flexibility \cite{long_REMD_GACC}
Despite the fact that GACC has been widely studied, it still remains challenging to effectively sample possible configurations and is a good system to test new methods\cite{Sponer2018}. For example, in a previous study, in order to get the converged structural populations, a multidimensional replica exchange molecular dynamics (M-REMD) simulation was performed with 192 replicas with around $1$ $\mu s$ of simulation time per replica, thus totalling almost 192 microseconds of all-atom simulations.\cite{long_REMD_GACC} Here we show that with DDPM, we can better estimate the free energy landscape using fewer computational resources, totalling only 12 microseconds of all-atom simulations adding all replicas.

We trained our DDPM model on 250 $ns$ REMD trajectories from 48 replicas with temperatures ranging from 277 K to 408 K (see Section MD simulations setup for details). In each frame of the trajectory, the structure of GACC is characterized by 6 dihedral angles for each nucleotide. In these REMD simulations, the sampling is not sufficient, especially for replicas at very low temperatures. Fig. \ref{fig:REMD_GACC} and Supplementary Fig. S4-S5 shows the free energy profiles of GACC projected on the $\gamma$ and $\zeta$ angles of A2 and C3 at different temperatures. We can see that expected high energy metastable state indicated by black arrows in Fig. \ref{fig:REMD_GACC} were not sampled in REMD. In contrast, for different target temperatures, DDPM successfully generates the ensemble of such high free energy states that were never even visited at low temperatures. Here as well, any spurious configurations can be seen through dots outside the free energy contours in Fig. \ref{fig:fes_GACC} with negligible Boltzmann weights.



\section*{Discussion}
We have presented a generative AI-based approach that combines physics from simulations performed at different temperatures to generate reliable new molecular configurations and accurate thermodynamic estimates at any arbitrary temperature even if no actual simulation was performed there. The central idea is to not work with the thermodynamic temperature $\beta$ of the system as a parameter set by the heat bath, but instead work with an effective temperature, calculated from the instantaneous kinetic energy of the system. This effective temperature shows non-trivial fluctuations for a finite size system, and on an average equals the thermodynamic temperature. Given sparse sampling from the high-dimensional space comprising configurational space coordinates and the effective temperature, we train a generative AI model that generates countless more samples of configurations at any temperature of interest.  While the idea is generally applicable, here we demonstrate its usefulness in the context of the widely used replica exchange molecular dynamics (REMD) framework to improve the sampling of REMD through a post-processing framework. We show how this significantly improves the quality of sampling at low temperatures and even generate samples in states where have not even been visited in the replicas, and at temperatures not considered in the ladder of replicas.  It is worth mentioning here that a recent application\cite{dibak2020temperature} of normalizing flows also attempts to enhance REMD sampling through somewhat similar ideas as ours. However in that work the machinery is used to directly affect the acceptance protocol in REMD while ours is a purely post-processing scheme.  Our work also bears parallel with the T-WHAM approach of Ref. \onlinecite{gallicchio2005temperature} which uses a similar joint probability in configuration and energy space after binning them. However, as shown in the Supplementary Fig. S9, numerically our approach is far superior as we do not compute weights involving exponentials of the full system's kinetic or potential energy, thereby avoiding well-known issues with the convergence of exponentially averaged quantities \cite{jarzynski2006rare}.

The generative AI framework of DDPM used here belongs to the broad class of flow-type methods, which have been shown to have the ability to generate samples from high dimensional space with many interdependent degrees of freedom. Compared with other flow type models such as normalizing flow\cite{NF, Boltzmann_generators}  that use deterministic functions to map from an easy-to-sample distribution to target distribution, the stochastic nature of DDPM avoids the restriction of preserving the topology of configuration space and thus allows the learning of significantly more complicated distributions. At the same time the design of the transition kernels in DDPM reduces the learning task to just learning means of Gaussian kernels. This makes the training easier compared to other methods\cite{stochastic_NF} while at the same time being able to learn more complicated transition kernels. 

We also believe that our generative AI model, while ``dreaming'' thermodynamically relevant structures at different temperatures, avoids the so-called hallucinations suffered by other generative AI models, i.e. we do not generate meaningless, unphysical structures with significant thermodynamic weights \cite{deep_learning_book}. We believe this is through the use of relatively simple transition kernels, which avoids overparameterization of the model, and through utilizing molecular basis functions instead of all-atom coordinates, which reduces the space that needs to be sampled. The issue of generating out-of-distribution samples that has been problematic in other methods attempting to generate molecular structure, such as the Boltzmann generator, is usually avoided by discarding the translational and rotational degrees of freedom and reweighting the samples.\cite{Boltzmann_generators, stochastic_NF} However, calculating the weight of each sample requires knowledge of all the coordinates of a system which may also become an issue when the system contains explicit water molecules; in such a configuration space the samples will again become sparse.  We finally point out that this work shows the possibility of learning generative models in the space of generic thermodynamic ensembles, by following the simple recipe that control parameters can also be viewed as fluctuating variables. As long as one is not in the thermodynamic limit -- something we do not have to worry about in molecular simulations -- this should be thus a practical and useful procedure for problems far beyond replica exchange molecular dynamics.

\subsection*{MD simulations setup}
\label{sec:setup}
\subsubsection*{AIB$_9$}
The simulation of AIB$_9$ was setup by following a previous study\cite{SPIB_AIB9}. The PDB file was taken from the authors with permission, and the simulations are done with the CHARMM36m all atom force field \cite{charmm36m} using TIP3P water molecules\cite{tip3p}, a Parrinello-Rahman barostat\cite{Parrinello_Rahman_dynamics}, and a Nose-Hoover thermostat\cite{hoover, nose--hoover} under the NPT ensemble. Simulations were performed using GROMACS 2016.\cite{2016gromacs} The structures of AIB$_9$ were saved every 0.2 ps and the dihedral angles were calculated using PLUMED 2.4\cite{plumed}. 

\subsubsection*{GACC}
The simulations of GACC were done with the AMBER ff12 all atom force field, using TIP3P water molecules\cite{tip3p}, a Parrinello-Rahman barostat\cite{Parrinello_Rahman_dynamics}, and a Bussi-Parrinello velocity rescaling thermostat\cite{Bussi_Parrinello_thermostat} under the NPT ensemble. The simulations were performed using GROMACS 2016\cite{2016gromacs}. The AMBER force filed was chosen because it exhibits more conformational variability compared with CHARMM force filed in a previous study\cite{Bergonzo2015}. The structures of GACC were also saved every 0.2 ps and the dihedral angles defined in Table S1 were calculated using PLUMED 2.4\cite{plumed}.

The PDB file of the GACC structure that served as the starting point for our simulation was generated using PyMOL. The initial structure was assumed to be at a temperature of 10K, and the systems energy was minimized with positional restraints of 25 kcal mol$^{-1} \AA^{-2}$ in a two step process - first the steepest descent algorithm was applied for 1000 steps followed by the conjugate gradient algorithm for another 1000 steps. Next, the system was equilibrated to 150K over 100ps under the NVT ensemble with positional restraints of 25 kcal mol$^{-1} \AA^{-2}$. Then, the GACC was equilibrated from 150K to 277K at 1 atm under the NPT ensemble for 100 ps with positional restraints of 5 kcal mol$^{-1} \AA^{-2}$. Finally, a long 5 ns equilibration was performed over 5ns at 298K under the NPT ensemble with positional restraints of 0.5 kcal mol$^{-1} \AA^{-2}$, after which the system was copied to 48 replicas, and each replica was equilibrated to its target temperature under the NPT ensemble with positional restraints of 0.5 kcal mol$^{-1} \AA^{-2}$. 

The replica temperatures were geometrically spaced temperatures ranging from 277K to 408K, with temperature increased by 1$\%$ for each replica. The attempt of exchanging configurations was made every 10 ps, which is determined by checking the time correlation function of the potential energy.

\subsection*{Network structure and hyperparameters}

The structure of the network is shown in Fig. \ref{fig:U-net}. It has a U-net structure\cite{u_net}, where the input is down-sampled by 4 residue blocks and then up-sampled by another 4 residue blocks. The diffusion process is divided into 1000 discrete steps. The network parameters are optimized by the Adam algorithm \cite{Adam} with a learning rate equal to $2\time 10^{-5}$. Exponential moving average (EMA) with a decrease rate 0.995 is used to stabilize the stochastic gradient descent.  

\showmatmethods{} 

\acknow{The authors would like to thank Alexander A. Alemi for initially giving us the idea of considering diffusion probabilistic models. We thank Shams Mehdi for helping set up the AIB$_9$ simulation. We also thank Zachary Smith, Ziyue Zou, Eric Beyerle, Bodhi Vani and Dedi Wang for proofreading the manuscript. This work was supported by the National Science Foundation, Grant No. CHE-2044165 and used XSEDE Bridges through allocation TG-CHE180053, which is supported by National Science Foundation grant number ACI-1548562. Y.W. would like to thank NCI-UMD Partnership for Integrative Cancer Research for financial support.}

\showacknow{} 


\begin{thebibliography}{10}

\bibitem{laidler1984development}
KJ Laidler, The development of the arrhenius equation.
\newblock {\em\protect\JournalTitle{Journal of chemical Education}}
  \textbf{61}, 494 (1984).

\bibitem{scalley1997protein}
ML Scalley, D Baker, Protein folding kinetics exhibit an arrhenius temperature
  dependence when corrected for the temperature dependence of protein
  stability.
\newblock {\em\protect\JournalTitle{Proceedings of the National Academy of
  Sciences}} \textbf{94}, 10636--10640 (1997).

\bibitem{chan1998protein}
HS Chan, KA Dill, Protein folding in the landscape perspective: Chevron plots
  and non-arrhenius kinetics.
\newblock {\em\protect\JournalTitle{Proteins: Structure, Function, and
  Bioinformatics}} \textbf{30}, 2--33 (1998).

\bibitem{DPM}
J Sohl-Dickstein, E Weiss, N Maheswaranathan, S Ganguli, Deep unsupervised
  learning using nonequilibrium thermodynamics in {\em International Conference
  on Machine Learning}.
\newblock (PMLR), pp. 2256--2265 (2015).

\bibitem{denoising}
J Ho, A Jain, P Abbeel, Denoising diffusion probabilistic models.
\newblock {\em\protect\JournalTitle{arXiv preprint arXiv:2006.11239}} (2020).

\bibitem{REMD}
Y Sugita, Y Okamoto, Replica-exchange molecular dynamics method for protein
  folding.
\newblock {\em\protect\JournalTitle{Chemical Physics Letters}} \textbf{314},
  141 -- 151 (1999).

\bibitem{hansmann1997parallel}
UH Hansmann, Parallel tempering algorithm for conformational studies of
  biological molecules.
\newblock {\em\protect\JournalTitle{Chemical Physics Letters}} \textbf{281},
  140--150 (1997).

\bibitem{abrams2014enhanced}
C Abrams, G Bussi, Enhanced sampling in molecular dynamics using metadynamics,
  replica-exchange, and temperature-acceleration.
\newblock {\em\protect\JournalTitle{Entropy}} \textbf{16}, 163--199 (2014).

\bibitem{abel2017advancing}
R Abel, L Wang, ED Harder, B Berne, RA Friesner, Advancing drug discovery
  through enhanced free energy calculations.
\newblock {\em\protect\JournalTitle{Accounts of chemical research}}
  \textbf{50}, 1625--1632 (2017).

\bibitem{REST}
P Liu, B Kim, RA Friesner, B Berne, Replica exchange with solute tempering: A
  method for sampling biological systems in explicit water.
\newblock {\em\protect\JournalTitle{Proceedings of the National Academy of
  Sciences}} \textbf{102}, 13749--13754 (2005).

\bibitem{REST2}
L Wang, RA Friesner, B Berne, Replica exchange with solute scaling: a more
  efficient version of replica exchange with solute tempering (rest2).
\newblock {\em\protect\JournalTitle{The Journal of Physical Chemistry B}}
  \textbf{115}, 9431--9438 (2011).

\bibitem{ballard2009replica}
AJ Ballard, C Jarzynski, Replica exchange with nonequilibrium switches.
\newblock {\em\protect\JournalTitle{Proceedings of the National Academy of
  Sciences}} \textbf{106}, 12224--12229 (2009).

\bibitem{trebst2006optimized}
S Trebst, M Troyer, UH Hansmann, Optimized parallel tempering simulations of
  proteins.
\newblock {\em\protect\JournalTitle{The Journal of chemical physics}}
  \textbf{124}, 174903 (2006).

\bibitem{nadler2007optimizing}
W Nadler, UH Hansmann, Optimizing replica exchange moves for molecular
  dynamics.
\newblock {\em\protect\JournalTitle{Physical Review E}} \textbf{76}, 057102
  (2007).

\bibitem{kim2010generalized}
J Kim, T Keyes, JE Straub, Generalized replica exchange method.
\newblock {\em\protect\JournalTitle{The Journal of chemical physics}}
  \textbf{132}, 224107 (2010).

\bibitem{chodera2011replica}
JD Chodera, MR Shirts, Replica exchange and expanded ensemble simulations as
  gibbs sampling: Simple improvements for enhanced mixing.
\newblock {\em\protect\JournalTitle{The Journal of chemical physics}}
  \textbf{135}, 194110 (2011).

\bibitem{gil2015enhanced}
A Gil-Ley, G Bussi, Enhanced conformational sampling using replica exchange
  with collective-variable tempering.
\newblock {\em\protect\JournalTitle{Journal of chemical theory and
  computation}} \textbf{11}, 1077--1085 (2015).

\bibitem{deep_learning_book}
I Goodfellow, Y Bengio, A Courville, {\em Deep Learning}.
\newblock (MIT Press), (2016) \url{http://www.deeplearningbook.org}.

\bibitem{gallicchio2005temperature}
E Gallicchio, M Andrec, AK Felts, RM Levy, Temperature weighted histogram
  analysis method, replica exchange, and transition paths.
\newblock {\em\protect\JournalTitle{The Journal of Physical Chemistry B}}
  \textbf{109}, 6722--6731 (2005).

\bibitem{u_net}
O Ronneberger, P Fischer, T Brox, U-net: Convolutional networks for biomedical
  image segmentation in {\em International Conference on Medical image
  computing and computer-assisted intervention}.
\newblock (Springer), pp. 234--241 (2015).

\bibitem{group_normalizng}
Y Wu, K He, Group normalization in {\em Proceedings of the European Conference
  on Computer Vision (ECCV)}.
\newblock (2018).

\bibitem{attention}
A Vaswani, et~al., Attention is all you need in {\em Advances in neural
  information processing systems}.
\newblock pp. 5998--6008 (year?).

\bibitem{vincent2011connection}
P Vincent, A connection between score matching and denoising autoencoders.
\newblock {\em\protect\JournalTitle{Neural computation}} \textbf{23},
  1661--1674 (2011).

\bibitem{tip3p}
WL Jorgensen, J Tirado-Rives, Potential energy functions for atomic-level
  simulations of water and organic and biomolecular systems.
\newblock {\em\protect\JournalTitle{Proceedings of the National Academy of
  Sciences}} \textbf{102}, 6665--6670 (2005).

\bibitem{huang2017charmm36m}
J Huang, et~al., Charmm36m: an improved force field for folded and
  intrinsically disordered proteins.
\newblock {\em\protect\JournalTitle{Nature methods}} \textbf{14}, 71--73
  (2017).

\bibitem{botan2007energy}
V Botan, et~al., Energy transport in peptide helices.
\newblock {\em\protect\JournalTitle{Proceedings of the National Academy of
  Sciences}} \textbf{104}, 12749--12754 (2007).

\bibitem{biswas2018metadynamics}
M Biswas, B Lickert, G Stock, Metadynamics enhanced markov modeling of protein
  dynamics.
\newblock {\em\protect\JournalTitle{The Journal of Physical Chemistry B}}
  \textbf{122}, 5508--5514 (2018).

\bibitem{SPIB_AIB9}
S Mehdi, D Wang, S Pant, P Tiwary, Accelerating all-atom simulations and
  gaining mechanistic understanding of biophysical systems through state
  predictive information bottleneck (2021).

\bibitem{Bannwarth2005}
S Bannwarth, A Gatignol, {HIV}-1 {TAR} {RNA}: The target of molecular
  interactions between the virus and its host.
\newblock {\em\protect\JournalTitle{Current {HIV} Research}} \textbf{3}, 61--71
  (2005).

\bibitem{SchulzeGahmen2018}
U Schulze-Gahmen, JH Hurley, Structural mechanism for {HIV}-1 {TAR} loop
  recognition by tat and the super elongation complex.
\newblock {\em\protect\JournalTitle{Proceedings of the National Academy of
  Sciences}} \textbf{115}, 12973--12978 (2018).

\bibitem{Ganser2019}
LR Ganser, ML Kelly, D Herschlag, HM Al-Hashimi, The roles of structural
  dynamics in the cellular functions of {RNAs}.
\newblock {\em\protect\JournalTitle{Nature Reviews Molecular Cell Biology}}
  \textbf{20}, 474--489 (2019).

\bibitem{Sponer2018}
J {\v{S}}poner, et~al., {RNA} structural dynamics as captured by molecular
  simulations: A comprehensive overview.
\newblock {\em\protect\JournalTitle{Chemical Reviews}} \textbf{118}, 4177--4338
  (2018).

\bibitem{long_REMD_GACC}
C Bergonzo, NM Henriksen, DR Roe, r Cheatham, Thomas~E, Highly sampled
  tetranucleotide and tetraloop motifs enable evaluation of common rna force
  fields.
\newblock {\em\protect\JournalTitle{RNA (New York, N.Y.)}} \textbf{21},
  1578--1590 (2015).

\bibitem{dibak2020temperature}
M Dibak, L Klein, F No{\'e}, Temperature-steerable flows.
\newblock {\em\protect\JournalTitle{arXiv preprint arXiv:2012.00429}} (2020).

\bibitem{jarzynski2006rare}
C Jarzynski, Rare events and the convergence of exponentially averaged work
  values.
\newblock {\em\protect\JournalTitle{Physical Review E}} \textbf{73}, 046105
  (2006).

\bibitem{NF}
D Rezende, S Mohamed, Variational inference with normalizing flows in {\em
  International conference on machine learning}.
\newblock (PMLR), pp. 1530--1538 (2015).

\bibitem{Boltzmann_generators}
F No{\'e}, S Olsson, J K{\"o}hler, H Wu, Boltzmann generators: Sampling
  equilibrium states of many-body systems with deep learning.
\newblock {\em\protect\JournalTitle{Science}} \textbf{365} (2019).

\bibitem{stochastic_NF}
H Wu, J K{\"o}hler, F No{\'e}, Stochastic normalizing flows.
\newblock {\em\protect\JournalTitle{arXiv preprint arXiv:2002.06707}} (2020).

\bibitem{charmm36m}
J Huang, et~al., Charmm36m: an improved force field for folded and
  intrinsically disordered proteins.
\newblock {\em\protect\JournalTitle{Nature methods}} \textbf{14}, 71--73
  (2017).

\bibitem{Parrinello_Rahman_dynamics}
M Parrinello, A Rahman, Crystal structure and pair potentials: A
  molecular-dynamics study.
\newblock {\em\protect\JournalTitle{Phys. Rev. Lett.}} \textbf{45}, 1196--1199
  (1980).

\bibitem{hoover}
WG Hoover, Canonical dynamics: Equilibrium phase-space distributions.
\newblock {\em\protect\JournalTitle{Physical review A}} \textbf{31}, 1695
  (1985).

\bibitem{nose--hoover}
DJ Evans, BL Holian, The nose--hoover thermostat.
\newblock {\em\protect\JournalTitle{The Journal of chemical physics}}
  \textbf{83}, 4069--4074 (1985).

\bibitem{2016gromacs}
M Abraham, B Hess, D van~der Spoel, E Lindahl, Gromacs reference manual,
  version 2016.4.
\newblock {\em\protect\JournalTitle{Department of Biophysical Chemistry,
  University of Groningen}} (2016).

\bibitem{plumed}
M Bonomi, Promoting transparency and reproducibility in enhanced molecular
  simulations.
\newblock {\em\protect\JournalTitle{Nature methods}} \textbf{16}, 670--673
  (2019).

\bibitem{Bussi_Parrinello_thermostat}
G Bussi, M Parrinello, Accurate sampling using langevin dynamics.
\newblock {\em\protect\JournalTitle{Phys. Rev. E}} \textbf{75}, 056707 (2007).

\bibitem{Bergonzo2015}
C Bergonzo, NM Henriksen, DR Roe, TE Cheatham, Highly sampled tetranucleotide
  and tetraloop motifs enable evaluation of common {RNA} force fields.
\newblock {\em\protect\JournalTitle{{RNA}}} \textbf{21}, 1578--1590 (2015).

\bibitem{Adam}
DP Kingma, J Ba, Adam: A method for stochastic optimization.
\newblock {\em\protect\JournalTitle{arXiv preprint arXiv:1412.6980}} (2014).

\end{thebibliography}
\end{document}